\documentclass[preprint,prl,superscriptaddress,amsmath,amssymb]{revtex4}
\usepackage{graphicx}
\usepackage{color}
\usepackage{units}
\usepackage{easymat}
\usepackage[normalem]{ulem}
\usepackage{color, soul}

\newcommand{\ket}[1]{\vert #1 \rangle}

\definecolor{lightblue}{rgb}{0.90,0.90,1.00}

\def\myddots{\mathinner{ \mkern1mu\raise5pt\hbox{.} \mkern2mu\raise2.5pt\hbox{.} \mkern3mu\raise0pt\vbox{\kern0pt\hbox{.}}\mkern1mu}}

\begin{document}
\title{Quantum random walks without walking}
\author{K. Manouchehri}
\author{J.B. Wang}
\email{wang@physics.uwa.edu.au} \affiliation{School of Physics, The
University of Western Australia}
\address{35 Stirling Hwy Crawley WA 6009
Australia}
\date{\today}

\begin{abstract}
Quantum random walks have received much interest due to their
non-intuitive dynamics, which may hold the key to a new generation of
quantum algorithms. What remains a major challenge is a physical
realization that is experimentally viable and not limited to special
connectivity criteria. We present a scheme for walking on arbitrarily
complex graphs, which can be realized using a variety of quantum
systems such as a BEC trapped inside an optical lattice. This scheme
is particularly elegant since the walker is not required to
physically step between the nodes; only flipping coins is sufficient.
\end{abstract}


\maketitle

Random walks have been employed in virtually every science related
discipline to model everyday phenomena such as the DNA synapsis
\cite{Sessionsa1997}, animals' foraging strategies
\cite{Benichou2005}, diffusion and mobility in materials
\cite{Trautt2006} and exchange rate forecast \cite{Kilian2003}. They
have also found algorithmic applications, for example, in solving
differential equations \cite{Hoshino1971}, quantum monte carlo for
solving the many body Schr\"{o}dinger equation \cite{Ceperley1986},
optimization \cite{Berg1993}, clustering and classification
\cite{Scholl2003}, fractal theory \cite{Anteneodo2007} or even
estimating the relative sizes of Google, MSN and Yahoo search engines
\cite{Bar-Yossef2006}.  Whilst the so called \emph{classical} random
walks have been successfully utilized in such a diverse range of
applications, \emph{quantum} random walks are expected to provide us
with a new paradigm for solving many practical problems more
efficiently \cite{Aharonov1993, Knight2003-1}. In fact quantum walks
have already inspired efficient algorithms with applications in
connectivity and graph theory \cite{Kempe2003, Douglas2008}, as well
as quantum search and element distinctness \cite{Shenvi2003,
Childs2004}, due to their non-intuitive and markedly different
properties including faster \emph{mixing} and \emph{hitting} times.

The question we address in this paper is how to physically implement
a quantum random walk in the laboratory. Over the last few years
there have been several proposals for such a physical implementation
using Nuclear Magnetic Resonance \cite{Ryan2005}, cavity QED
\cite{Agarwal2005}, ion traps \cite{Travaglione2002}, classical and
quantum optics \cite{Knight2003-1, Zhang2007}, optical lattice and
microtraps \cite{Joo2006, Eckert2005} as well as quantum dots
\cite{Manouchehri2008-1, Solenov2006}. None of the existing proposals
however consider quantum random walks on general graphs, with the
majority describing only a one-dimensional implementation. This is
while from an application point of view most useful algorithms would
involve traversing graphs with arbitrarily complex structures.

In this paper, we present a scheme which considerably simplifies the
evolution of the quantum walk on a general undirected graph. We then
describe a systematic procedure capable of performing this quantum
walk on a variety of existing as well as prospective quantum
computing platforms. Finally we present an example of one such
implementation, using a Bose-Einstein condensate (BEC) of $^{87}$Rb
atoms trapped inside a 2D optical lattice \cite{Morsch2006}.

\begin{figure}[h]
    \text{\raisebox{3.4cm}{(a)}}\includegraphics[width=4cm, bb=0 0 247 220,clip]{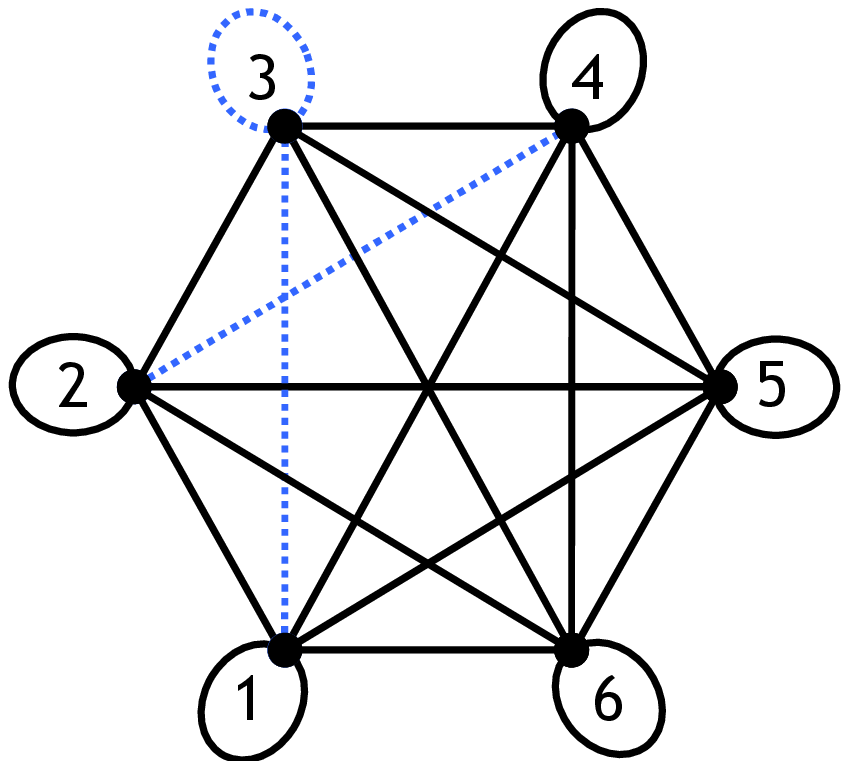}
    \text{\raisebox{3.4cm}{(b)}}\includegraphics[width=4cm, bb=0 0 470 500,clip]{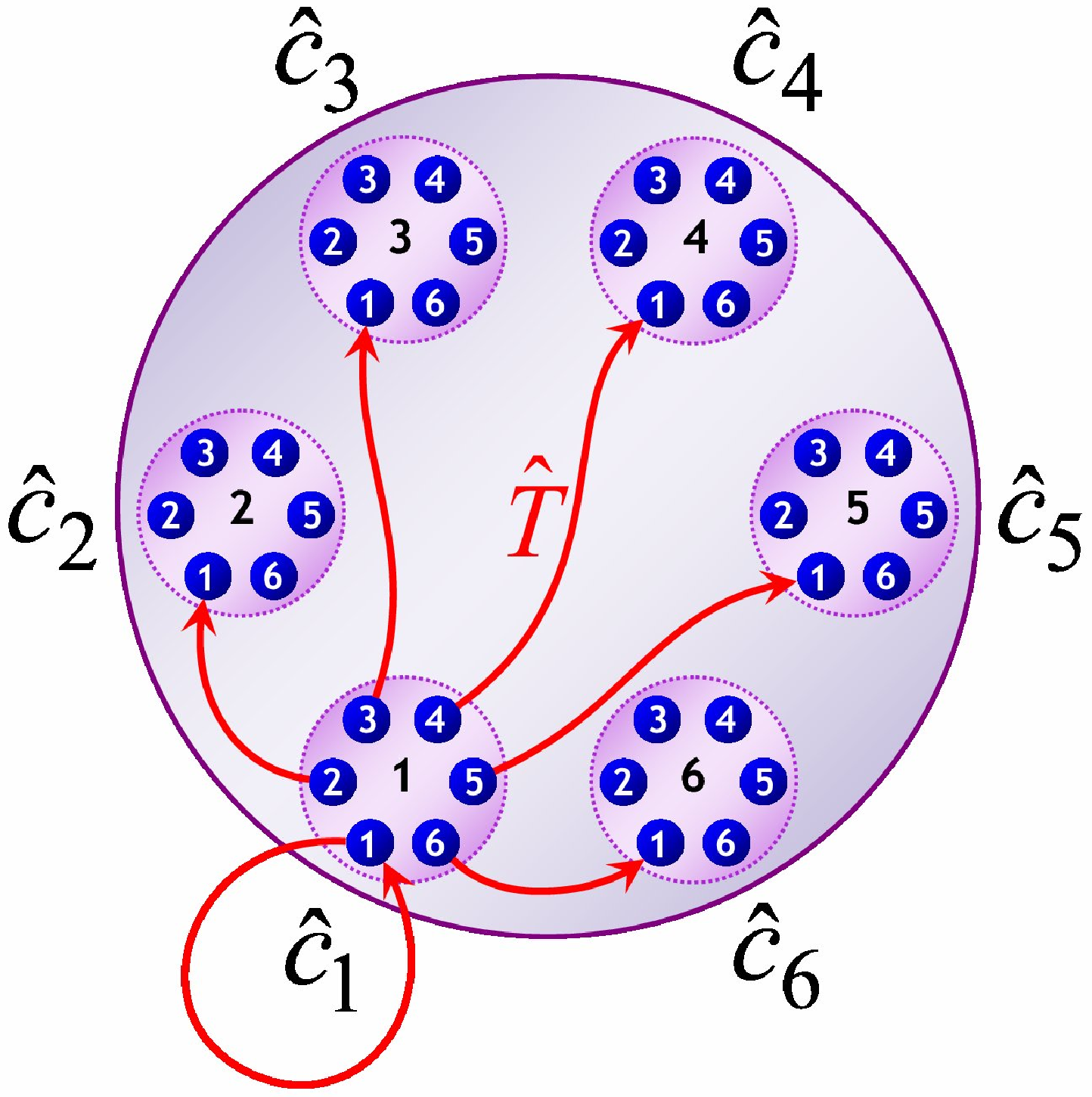}
    \text{\raisebox{3.4cm}{(c)}}\includegraphics[width=5cm, bb=0 0 390 380,clip]{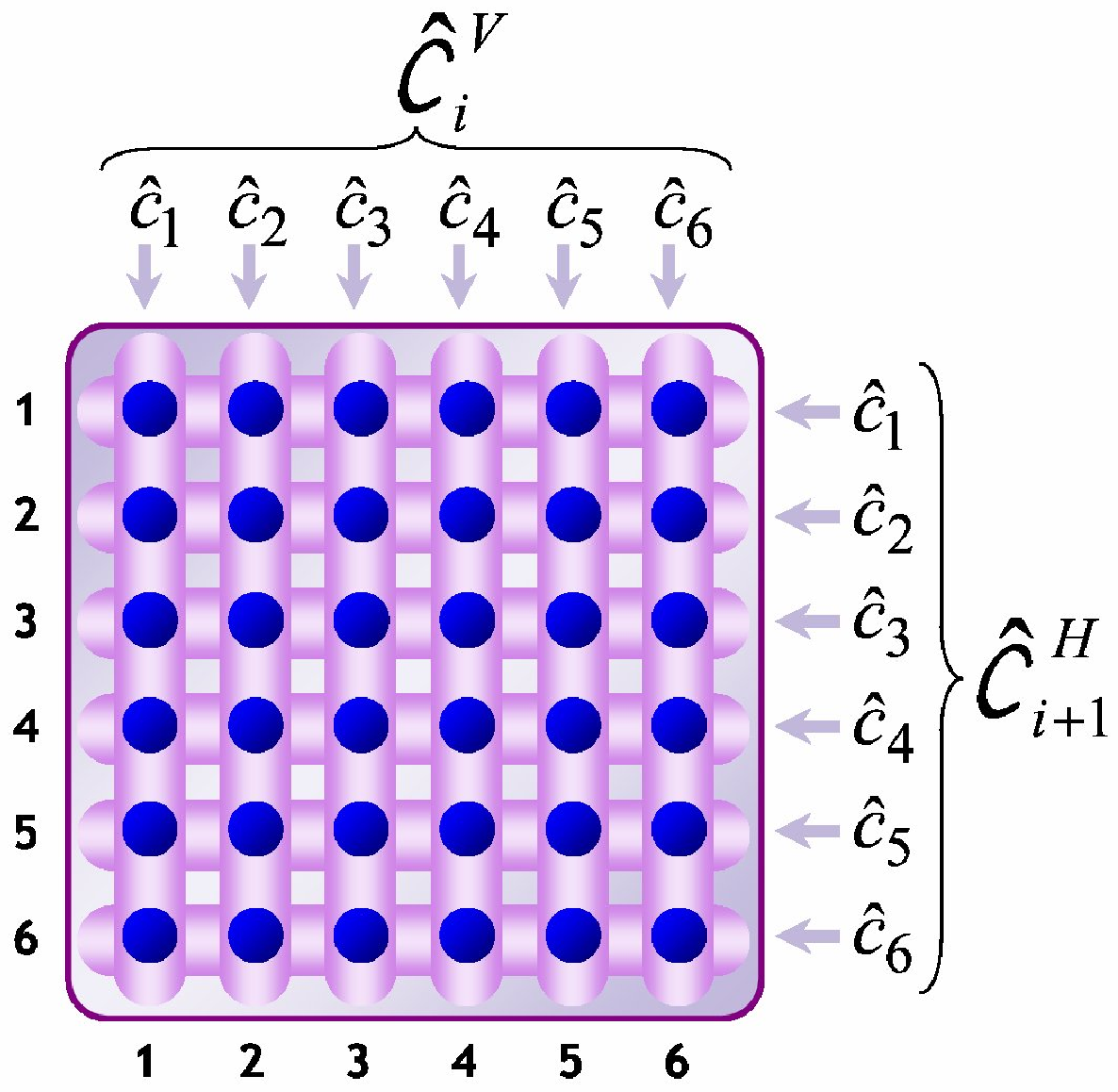}
    \caption{(a) A complete 6-graph. Any generalized graph can be constructed by removing edges (dotted
	lines) from the complete graph; (b) Quantum walk Hilbert space and a particular mapping
    $\hat{\mathcal{T}}\ket{j,k}\longrightarrow\ket{k,j}$; (c)
    $\hat{\mathcal{T}}$ is replaced by alternating the direction in which $\hat{\mathcal{C}}$
    is applied in successive steps of the walk. }
    \label{fig.coin-operation}
\end{figure}

First we consider a complete graph with all possible connections
between the $\mathcal{N}$ nodes including self loops
(Fig.~\ref{fig.coin-operation}a). Here the walker requires an
$\mathcal{N}$-sided coin for moving from one node to $\mathcal{N}$
other nodes. The complete state of the walker is therefore described
by $\ket{\psi} = \sum_{j=1}^\mathcal{N} \sum_{k=1}^\mathcal{N}
\mathcal{A}_{j,k}\ket{j,k}$, where $\mathcal{A}_{j,k}$ are complex
amplitudes, $\ket{j}$ and $\ket{k}$ represent the node and coin
states respectively. A quantum coin flip corresponds to a unitary
rotation of the coin states at every node $j$ using an
$\mathcal{N}\times\mathcal{N}$ matrix $\hat{c}_j$ also known as the
coin operator. The coin operation is followed by the walker stepping
from node $j$ simultaneously to all other nodes on the graph using a
conditional translation operator $\hat{T}$ such that
$\hat{T}\ket{j,k}\longrightarrow\ket{j',k'}$, where $j$ and $j'$
label the two nodes at the end of an edge $e_{jj'}$
\cite{Kendon2005}. The quantum walk evolves via repeated applications
of the coin followed by the translation operator. More explicitly, we
have $\ket{\psi_n} = \hat{T}_n~\hat{\mathcal{C}}_n
\ldots\hat{T}_2~\hat{\mathcal{C}}_2~\hat{T}_1~\hat{\mathcal{C}}_1 ~
\ket{\psi_0}$, where $\ket{\psi_0}$ is the initial state of the
walker, $\ket{\psi_n}$ is its state after $n$ steps,
$\hat{\mathcal{C}}_i$ and $\hat{T}_i$ are the coin and translation
operators at the $i$th step, and $\hat{\mathcal{C}}$ incorporates the
individual coin operators $\hat{c}_1 \ldots \hat{c}_\mathcal{N}$
which simultaneously act on all the nodes. The operators $\hat{c}$
can in principle invoke different rotations at each node $j$, but are
often uniformly set to be the Hadamard rotation.

In traversing the edge $e_{jj'}$, we define
$\hat{\mathcal{T}}\ket{j,k}\longrightarrow\ket{k,j}$
 (Fig.~\ref{fig.coin-operation}b).
Without undue loss of generality, this choice of translation operator
has the unique advantage of being independent of graph connectivity,
and thus enabling a quantum walk to be systematically implemented on
any arbitrary graph. Upon visualizing the Hilbert space of the walk
as an $\mathcal{N}\times\mathcal{N}$ square array $\mathcal{H}$ with
entries $h_{jk}$ representing the states $\ket{j, k}$, the
application of the translation operator $\hat{\mathcal{T}}$ to the
state space of the walk simply becomes equivalent to a transposition
of the array elements. Let us now consider the first few steps in the
evolution of a quantum walk. Applying $\hat{\mathcal{C}}_1$ to the
state space of the walk involves performing $\mathcal{N}$
simultaneous unitary transformations $\hat{c}_j$, each on the coin
states of the node corresponding to the $j$th row. This leads to a
natural grouping of the states along the rows of $\mathcal{H}$ and we
employ the relabeled operator $\hat{\mathcal{C}}_1^H$ to highlight
that it operates on \emph{horizontally} grouped states
(Fig.~\ref{fig.coin-operation}c). What is particularly convenient now
is that instead of transposing $\mathcal{H}$ due to the action of
$\hat{\mathcal{T}}_1$ we can simply transpose the application of the
next coin operator $\hat{\mathcal{C}}_2$. By transposing
$\hat{\mathcal{C}}_2$ we mean regrouping the states, this time along
the columns of $\mathcal{H}$, and performing $\mathcal{N}$
simultaneous unitary transformations $\hat{c}_j$, each on the states
of the $j$th column. As before we employ the relabeled operator
$\hat{\mathcal{C}}_2^V$ to highlight that it operates on
\emph{vertically} grouped states.

In the above formulation, the effect of the translation operator
$\hat{\mathcal{T}}$ is implicit in the regrouping of states and does
not appear in the expression governing the evolution of the walk,
which can now be written as $\ket{\psi_n} = \hat{\mathcal{C}}_n^V
\hat{\mathcal{C}}_{n-1}^H \ldots
\hat{\mathcal{C}}_2^V~\hat{\mathcal{C}}_1^H ~ \ket{\psi_0}$, halving
the number of required operations. It is in this sense that we have
qualified this process as a ``quantum random walk \emph{without
walking}''; the walker is not required to physically step between the
nodes, only flipping the coin is sufficient. As we will see, removing
the quantum walk's dependence on the translation operator
$\hat{\mathcal{T}}$ greatly facilitates its physical implementation.

We now construct our intended graph $\mathcal{G}$ by simply removing
all the unwanted edges (dotted lines in
Fig.~\ref{fig.coin-operation}a) from its complete counterpart
$\mathcal{G}_{\max}$. In turn this has the effect of removing some of
the states from the Hilbert space $\mathcal{H}$ (dotted circles in
Fig.~\ref{fig.coin-operation}c). Removing the edge $e_{jj'}$ for
example, corresponds to removing two states $\ket{j,j'}$ and
$\ket{j',j}$. In our approach however, instead of removing these
unwanted states from $\mathcal{H}$, we simply isolate them from
interaction with other states by appropriately designing the coin
operators $\hat{c}_1\ldots\hat{c}_\mathcal{N}$. Taking
$\hat{\mathcal{C}}^H$ as an example, matrix $\hat{c}_j^H$ performs a
unitary transformation on the $j$th row of $\mathcal{H}$. Hence to
isolate the state $\ket{j,k}$ we obtain a modified coin matrix whose
column elements $c_{1k}\ldots c_{\mathcal{N}k}$ and row elements
$c_{k1}\ldots c_{k\mathcal{N}}$ are all set to zero except for
$c_{kk}$ which is 1. Using this modified coin matrix guarantees that
if initially the walker has no amplitude in state $\ket{j,k}$, this
state will remain unpopulated throughout the evolution of the walk.

It is clear from the preceding discussion that a physical
implementation of this walk requires two basic properties commonly
found in a variety of systems proposed for traditional quantum
computing: (a) $\mathcal{N}^2$ basis states arranged in a square
array formation and (b) implementing the operators $\hat{c}_j^H ~
(\hat{c}_{j'}^V)$, which at once perform an $\mathcal{N}$-state
unitary rotation on all the amplitudes in row $j$ (column $j'$) of
the 2D state space. Such a mechanism can indeed be efficiently
constructed if the system is capable of performing pairwise unitary
operations on \emph{non-neighboring} states similar to those
demonstrated in \cite{Mandel2003-1, Mandel2003-2,  Lee2007,
Majer2007} and discussed in \cite{Calarco2004, Pedersen2008} and
references therein. The key to our implementation is a Cosine Sine
(CS) decomposition \cite{Sutton2009} which effectively takes the
single unitary operator $\hat{c}_j^H ~ (\hat{c}_{j'}^V)$ and replaces
it with a series of pairwise operators which we know how to
implement. One requirement of this implementation is that
$\mathcal{N} = 2^{N}$ for some integer $N$, which can introduce some
redundancy in the the Hilbert space of the quantum walk, but only
adds a linear overhead. Considering the wave function along row $j$,
we represent the operator $\hat{c}_j^H$ as an $\mathcal{N}\times
\mathcal{N}$ unitary matrix acting on a vector $\mathbf{A}_j^H =
\left(\alpha_1 \cdots \alpha_{\mathcal{N}} \right)$ of amplitudes in
row $j$. Performing $N - 1$ recursive CS decompositions on
$\hat{c}_j^H$ we obtain
\begin{equation}\label{eqn.cs-product}
    \hat{c}_j^H = \prod_{i = 1}^{\mathcal{N} - 1} \mathcal{U}_i(d_i), \text{~where~}
    \mathcal{U}_i(d_i) =
        \left(\begin{MAT}{ccc}
                u_{i,1} & & \\
                & u_{i,2} & \\
                & &  \myddots \\
        \end{MAT} \right)
\end{equation}
and $u_{i,k}$ represent $d_i \times d_i$ square blocks along the
$\mathcal{U}_i$ diagonal with $k =1, 2 \cdots \mathcal{N}/d_i$. Block
dimensions can vary for each $\mathcal{U}_i$ with values restricted
to $d_i=2, 4, 8 \cdots \mathcal{N}/2$. For $d_i=2$, blocks $u_{i,k}$
represent general $2\times 2$ unitary matrices, but for $d_i>2$ they
assume the special form
\begin{equation}\label{eqn.CS-matrix}
   u_{i,k} = \left(\begin{MAT}(b,0.05pt,0.05pt){rrrrrr}
           \myddots & & & \myddots & & \\
           & c_r & & & ~s_r & \\
           & & \myddots & & & \myddots \\
           \myddots & & & \myddots & & \\
           & - s_r & & & c_r & \\
           & & \myddots & & & \myddots
          \addpath{(1,5,.)rrrrddddlllluuuu}
          \addpath{(0,3,3)rrrrrr}
          \addpath{(3,0,3)uuuuuu} \\
       \end{MAT}
     \right)_{i,k},
\end{equation}
where each quadrant is diagonal with respective entries $c_r$ and
$s_r$ corresponding to $\cos(\phi_r)$ and $\sin(\phi_r)$ for some
angle $\phi_r$ and $r=1, 2 \cdots d/2$. The action of each matrix
$\mathcal{U}_i(d_i)$ on the vector $\mathbf{A}_j^H$ can now be
directly implemented using pairwise interactions. Upon a closer
examination of $u_{i,k}$ in Eq. \ref{eqn.CS-matrix} we find that each
$c_r s_r$ square block (dotted) performs a pairwise unitary
transformation $\overline{u}_{i,k,r}$ on the amplitudes
$\alpha_{(k-1)d+r}$ and $\alpha_{(k-1)d+r+d/2}$, which are
non-neighboring for $d > 2$. Hence the rotation $\mathcal{U}_i(d_i)$
can be applied at once by simultaneously activating pairwise
interactions between all states in the range
$\ket{j,kd-d+1}\ldots\ket{j, kd-d/2}$ and their corresponding
counterparts $\ket{j,kd-d/2+1}\ldots\ket{j,kd}$ for all $k$. Note
that conveniently, all interacting pairs of states have the same
interval $d/2$ which greatly facilitates the design of a physical
implementation.

In the following we describe one such physical implementation using a
BEC trapped in a 2D optical lattice \cite{Jaksch2004}, where states
$\ket{j,k}$ of the walk are encoded using the individual trapping
sites and the BEC wave function acts as the quantum walker with some
initial distribution throughout the lattice sites. The system is
driven into a Mott insulator phase \cite{Spielman2007} thereby
suppressing the tunneling between neighboring lattice sites. A series
of specially tailored control laser operations are then introduced to
address, manipulate and interact the BEC wave packets in individual
sites, in a way that corresponds exactly to the action of the
operators $\hat{c}_j^H ~ (\hat{c}_{j'}^V)$ along the lattice rows
(columns). Although the control laser wavelength and the lattice
period $\lambda_{\text{lattice}}$ are comparable in size, problems
associated with unwanted interactions of the control laser with
neighboring sites can be circumvented by adopting techniques such as
those detailed in \cite{Cho2008, Gorshkov2008} or more readily by
choosing every 2nd, 3rd or $\ell$th lattice site to represent the
walk states. From an application point of view one would commonly
start with the BEC entirely localized in one site or uniformly loaded
into every $\ell$th site using pattern loading \cite{Peil2003} or by
employing a recently developed imaging and manipulation technique
based on scanning electron microscopy \cite{Wurtz2009, Gericke2008}.
The design of all subsequent control operations ensures that the
initially empty intermediate sites would, in principle, remain
unpopulated throughout the walk. In practice however the spatial
separation $\ell$ also acts as a buffer zone to contain any spiling
of the BEC out of its confinement lattice-site due to unavoidable
experimental imperfections.

\begin{figure}[h]
    \text{\raisebox{4.5cm}{   }}\includegraphics[width=7.5cm, bb=0 0 345 300,clip]{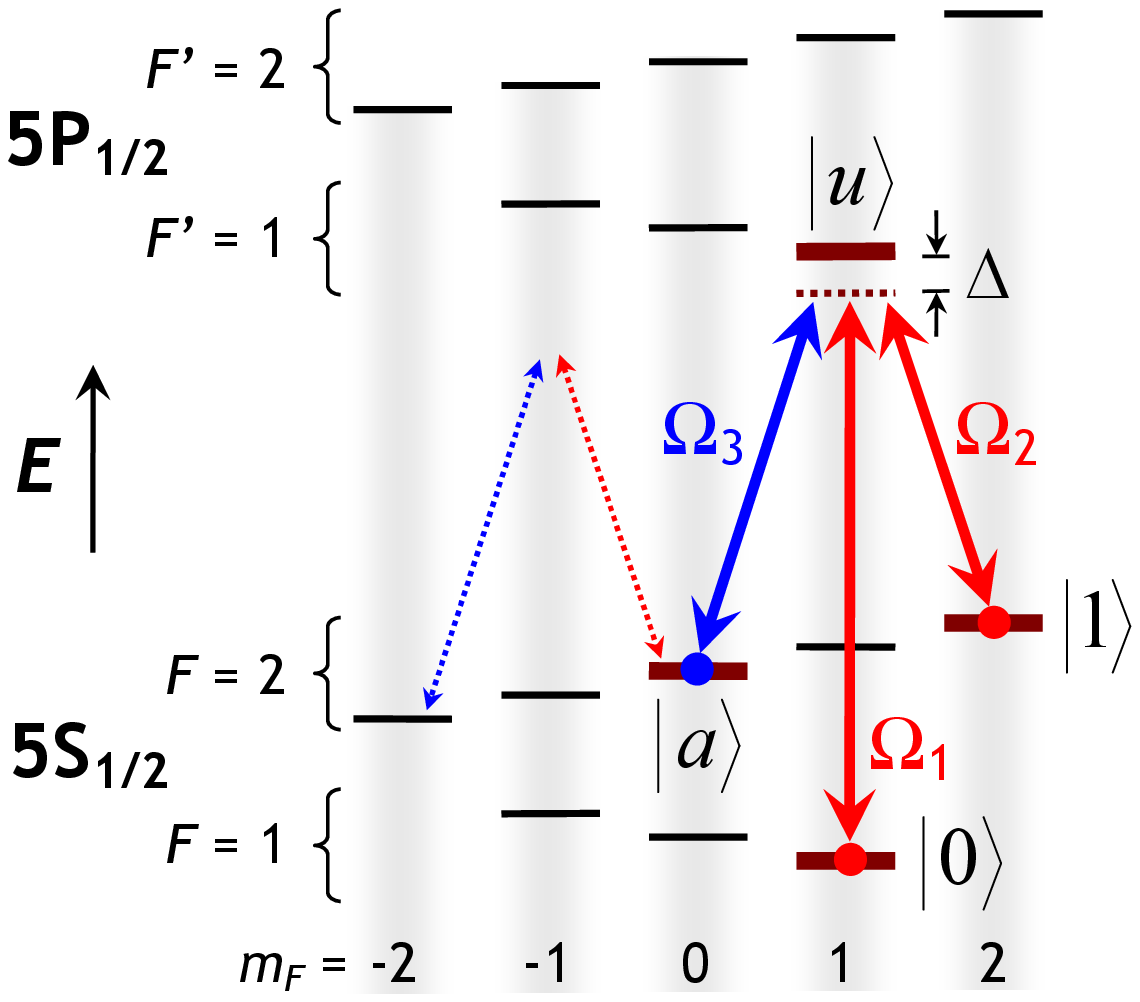}
    \text{\raisebox{4.5cm}{  }}\includegraphics[width=7.5cm, bb=0 0 355 260,clip]{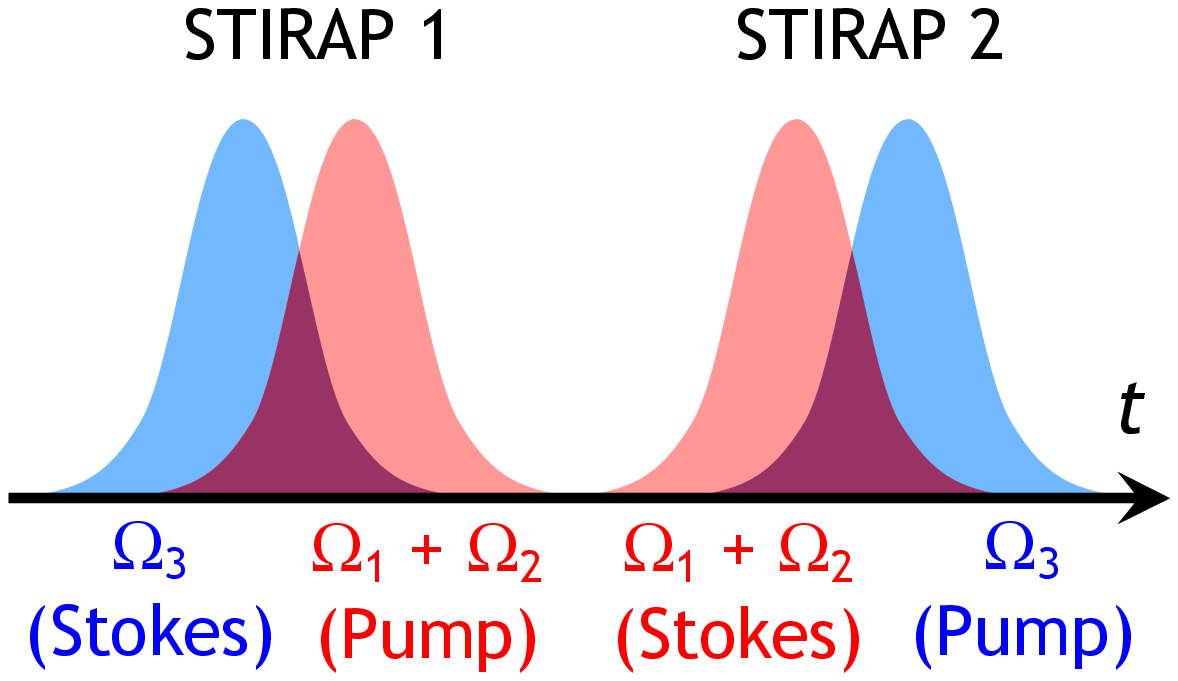}
    \caption{Schematic diagram of a three-photon STIRAP operation in a $^{87}
	$Rb atom. Internal levels $\ket{0}$, $\ket{1}$, $\ket{a}$ and $\ket{u}$ are
	coupled by three laser pulses with frequencies $\Omega_1$, $\Omega_2$ and $
	\Omega_3$ and polarizations that are linear, left circular $\sigma_{-}$ and
	right circular $\sigma_{+}$ respectively.}
    \label{fig.stirap}
\end{figure}

To manipulate the trapped BEC wave packet at a given lattice site we
propose performing arbitrary unitary transformations on the internal
states $\ket{0} \equiv \ket{F=1,m_F=1}$ and $\ket{1} \equiv
\ket{F=2,m_F=2}$ of the BEC with the aid of a pair of three-photon
Stimulated Raman Adiabatic Passage (STIRAP) operations
\cite{Kis2002}. Each STIRAP requires the use of three control lasers
(with wavelengths $\sim\lambda_{\text{lattice}}$) applied in the
counter intuitive order to transfer the atomic population in states
$\ket{0}$ and $\ket{1}$, to and from an auxiliary state $\ket{a}
\equiv \ket{F=2,m_F=0}$, via an intermediate upper state $\ket{u}
\equiv \ket{F'=1,m_F=1}$ that does not get populated during the
transfer (Fig.~\ref{fig.stirap}). The two-photon $\Lambda$ STIRAP
$\ket{1}\longleftrightarrow\ket{u}\longleftrightarrow\ket{a}$ has
already been experimentally demonstrated using circularly polarized
lasers and a magnetic field to lift the degeneracy in the sub-levels
$m_F$ \cite{Wright2008}. Our proposal simply extends this
implementation through the addition of a third linearly polarized
laser to facilitate $\ket{0}\longleftrightarrow\ket{u}$.

For performing a unitary transformation of BEC amplitudes in a pair
of lattice sites, we utilize a scheme for the spin(state)-dependent
transport of neutral atoms in an optical lattice \cite{Mandel2003-1,
Mandel2003-2}. By setting the wavelength $\lambda_{\text{lattice}} =
785$nm, internal states $\ket{0}$ and $\ket{1}$ experience different
corresponding dipole potentials $\mathcal{V}_0(x, \theta) =
\frac{1}{4}V_+(x, \theta) + \frac{3}{4} V_-(x, \theta)$ and
$\mathcal{V}_1(x, \theta) = V_+(x, \theta)$, where $V_\pm(x, \theta)
= V_{\max} \cos^2(\widetilde{k} x \pm \theta / 2)$, $\widetilde{k} =
2\pi /\lambda_{\text{lattice}}$ is the wave vector of the laser light
propagating in the $x$ direction, and $\theta$ is the relative
polarization angle between the pair of counter-propagating lasers.
Hence for an atom in the superposition state $\alpha \ket{0} + \beta
\ket{1}$, increasing the polarization angle $\theta$ will lead to a
split in the spatial wave packet of the atom as it perceives a
relative motion between the two potentials, resembling that of a pair
of conveyor belts moving in opposite directions, each carrying one of
the components $\alpha$ and $\beta$. The relative displacement is
given by $\Delta x = \theta \lambda_{\text{lattice}} / 2\pi$.

\begin{figure}[h]
    \text{\raisebox{4cm}{(a) }}\includegraphics[width=11cm, bb=0 0 530 222,clip]{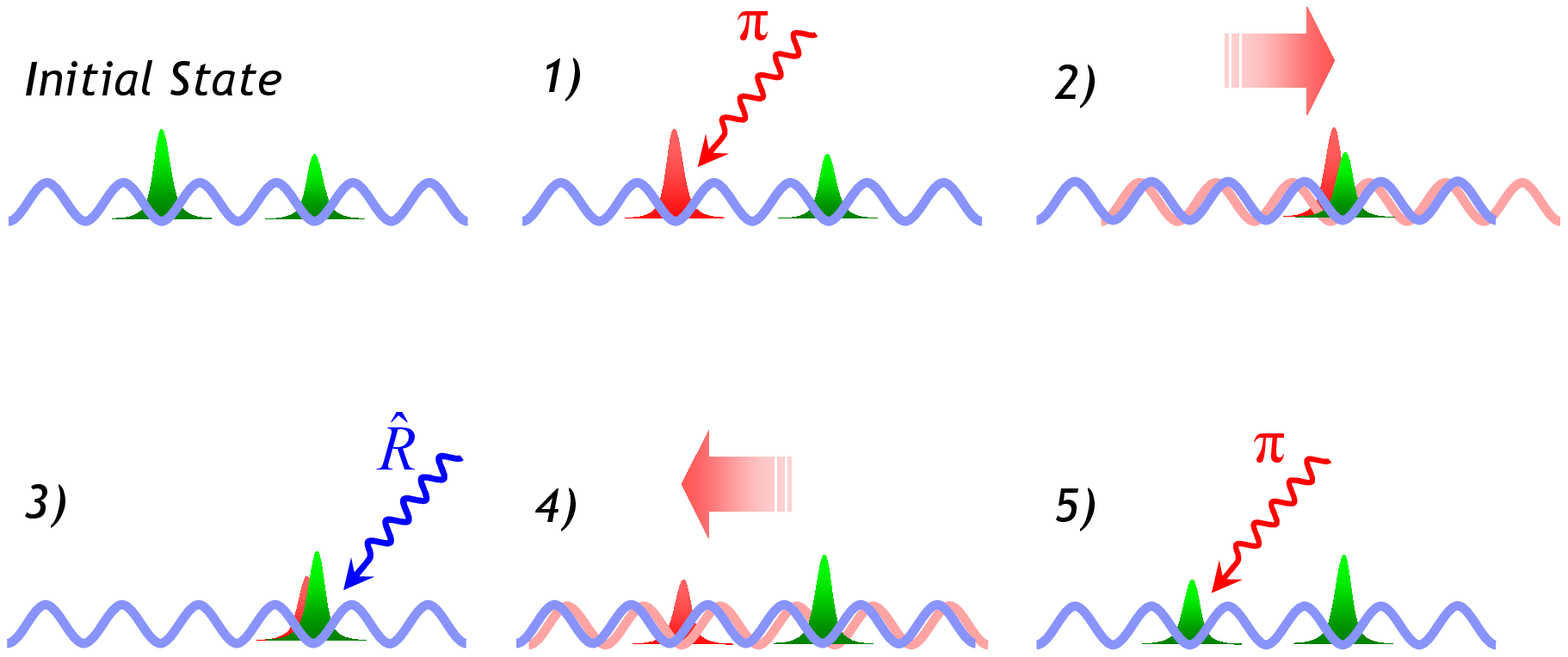}
    \text{\raisebox{8.2cm}{(b) }}\includegraphics[width=7.5cm, bb=0 0 440 550,clip]{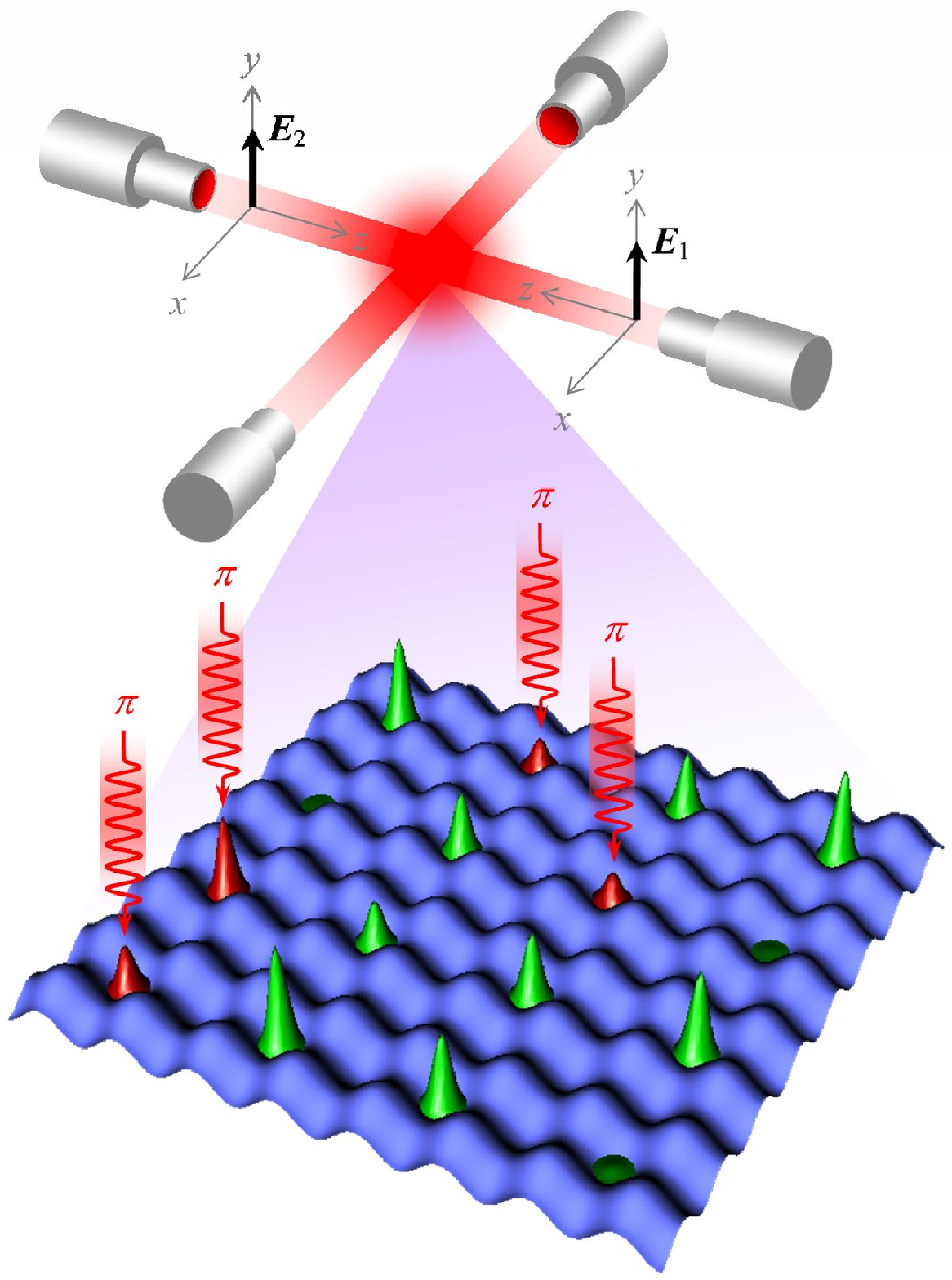}
    \text{\raisebox{8.2cm}{ }}\includegraphics[width=7.5cm, bb=0 0 440 550,clip]{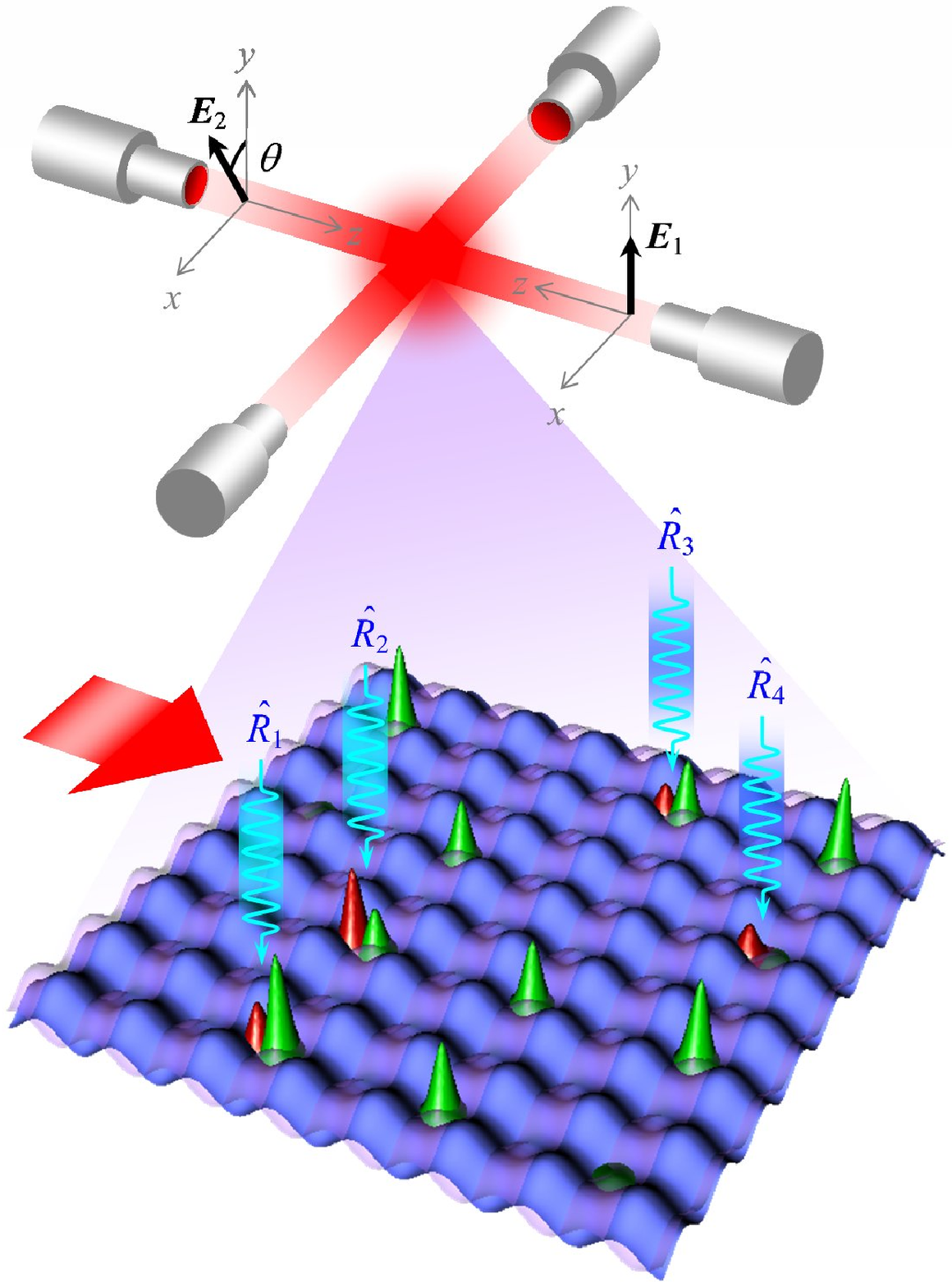}
    \caption{(a) Steps for applying a unitary transformation to BEC amplitudes
	trapped in a pair of non-neighboring optical lattice sites;
    (b) The first three steps on a 2D optical lattice with BEC site separation $\ell=2$.}
    \label{fig.interaction-procedure}
\end{figure}

Let us take a BEC initially prepared in the internal state $\ket{0}$
and distributed between two lattice sites $\ket{j,k}$ and
$\ket{j,k'}$ such that $\ket{\psi_0}=\alpha_k\ket{j,k}\otimes\ket{0}
+ \alpha_{k'}\ket{j,k'}\otimes\ket{0}$. We can now manipulate the
amplitudes $\alpha_k$ and $\alpha_{k'}$ according to any desired
unitary transformation in five steps depicted in
Fig.~\ref{fig.interaction-procedure}a. (1) Using the three-photon
STIRAP we apply a $\pi$-rotation to the BEC at $\ket{j,k}$ which
transfers it entirely to the internal state $\ket{1}$ and the new
state of the system becomes $\ket{\psi_1} =
\alpha_k\ket{j,k}\otimes\ket{1} +
\alpha_{k'}\ket{j,k'}\otimes\ket{0}$. (2) Making use of the
spin(state)-dependant transport, we increase the polarization angle
by $\theta = 2 \ell (k - k') \pi / \lambda_{\text{lattice}}$ causing
the two wave packets to fully overlap at $\ket{j,k'}$ (selected as
the stationary reference frame) and hence
$\ket{\psi_2}=\ket{j,k'}\otimes\left(\alpha_k\ket{1} +
\alpha_{k'}\ket{0}\right)$. (3) Using another three-photon STIRAP we
perform an arbitrary unitary rotation $\hat{R}$, this time at
$\ket{j,k'}$, such that
$\ket{\psi_3}=\ket{j,k'}\otimes\left(\tilde{\alpha}_k\ket{1} +
\tilde{\alpha}_{k'}\ket{0}\right)$. (4) Reversing the change in the
polarization angle we transport the new BEC amplitudes
$\tilde{\alpha}_k$ and $\tilde{\alpha}_{k'}$ back to their original
sites, i.e. $\ket{\psi_4}=\tilde{\alpha}_k\ket{j,k}\otimes\ket{1} +
\tilde{\alpha}_{k'}\ket{j,k'}\otimes\ket{0}$. (5) Finally performing
another $\pi$-rotation on the state $\ket{j,k}$ we transfer the BEC
back to the internal state $\ket{0}$ producing the desired outcome
$\ket{\psi_5}=\tilde{\alpha}_k\ket{j,k}\otimes\ket{0} +
\tilde{\alpha}_{k'}\ket{j,k'}\otimes\ket{0}$. Note that internal
states $\ket{0}$ and $\ket{1}$ are only used to facilitate the
pair-wise interactions and both BEC wave packets will be in their
internal ground state $\ket{0}$ before and after they interact.

This scheme can be readily extended to simultaneously activate all
the pair-wise interactions required for performing the unitary
rotations in Eq. \ref{eqn.cs-product}. We emphasize that all the
$\hat{c}_j^H ~ (\hat{c}_{j'}^V)$ operations along the rows (columns)
of the optical lattice are performed concurrently, since the
structure of the CS decomposition (Eq. \ref{eqn.CS-matrix}) is
identical for all coin operators and changing the polarization angle
$\theta$ triggers the same spin (state)-dependent transport across
the entire optical lattice. The effect of using different coin
operators for each node appears in step (3), where the control STIRAP
can perform different unitary rotations at various lattice sites. At
the conclusion of the walk, BEC densities throughout the lattice can
be determined via scanning electron microscopy
\cite{Wurtz2009,Gericke2008} or spin-selective absorption imaging
\cite{Greiner2003}, although the latter requires repeated runs of the
experiment for each node density measurement. The corresponding
quantum walk distribution is then derived by integrating the BEC
amplitudes over an area $\ell \lambda_{\text{lattice}} \times \ell
\lambda_{\text{lattice}}$ centered around the key lattice sites. This
will effectively include in the distribution, any residual amplitudes
in the neighboring intermediate sites, which are nonetheless
substantially lower than the amplitudes in key lattice sites and
would therefore have a minimal effect on the final result.

The proposed quantum walk scheme offers a polynomial speedup over an
equivalent quantum circuit implementation, highlighting the expected
trade off between resource and time scalability. A quantum circuit
can in principle represent the walk's Hilbert space using
$m=\log_2(\mathcal{N}^2)$ entangled qubits, which is by far more
resource efficient. Then, implementing a generalized $\mathcal{N}^2
\times \mathcal{N}^2$ unitary operator
$\hat{\mathcal{T}}_i~\hat{\mathcal{C}}_i$ for each step of the
quantum walk amounts to performing a $m$-qubit gate operation that
can be realized with around $4^m$ CNOT gates \cite{Vartiainen2004}.
Since the quantum circuit can perform at most $m/2$ simultaneous CNOT
operations at any one time, each step of the quantum walk requires at
least $(4^m)/(\nicefrac{m}{2})= 2 \mathcal{N}^4/
\log_2(\mathcal{N}^2)$ operational stages. This is compared to only
$\mathcal{N}-1$ operational stages needed for implementing Eq.
\ref{eqn.cs-product}.

Spin(state)-dependant BEC systems have also been considered as
serious contenders for building a quantum computer \cite{Monroe2002}.
This is despite the acute sensitivity of the BEC internal states
$\ket{0}$ and $\ket{1}$ to the external magnetic-field environment,
leading to phase decoherence times that are presently in the order of
a few $ms$ \cite{Lee2007}. Nonetheless, comparing this with a
single-site transport time ($\sim 50 \mu s$) \cite{Mandel2003-1,
Mandel2003-2} and STIRAP pulse durations ($\sim 60 \mu s$)
\cite{Wright2008}, and also noting the successful realization of
spin(state)-dependant BEC transport for up to 7 sites reported in
\cite{Mandel2003-1}, a ``proof of principle'' implementation (i.e.
the first few steps of the walk on an arbitrary graph with a few
nodes) should indeed be possible, utilizing the existing experimental
techniques. Since our proposed implementation scheme is in fact not
inherently bound to any one physical system, naturally as this and
other prospective quantum computing hardware grow in scale and
fidelity of operations, so will the complexity of graphs on which the
quantum walk can be performed.

\bibliographystyle{apsrev}
\bibliography{qrw-on-graphs}

\clearpage

\section{Figures}

\end{document}